\documentclass[twocolumn,showpacs,prl,amsmath,amssymb]{revtex4}
\usepackage{graphicx}
\newcommand{\be}{\begin{equation}}
\newcommand{\ee}{\end{equation}}
\newcommand{\beq}{\begin{eqnarray*}}
\newcommand{\eeq}{\end{eqnarray*}}

\def\o{\omega}

\begin{document}

\title{The Raman coupling function in disordered solids:
a light and neutron scattering study on glasses of different
fragility}

\author{
        A. Fontana$^{1,4}$, F. Rossi$^{1,4}$, G. Viliani$^{1,4}$,
        S. Caponi$^{1,4}$, and E. Fabiani$^{3}$
        }

\affiliation{
            $^1$Dipartimento di Fisica, Universit\`a di Trento, I-38050 Povo Trento, Italy\\
            $^2$IFN-CNR, via Sommarive 18, I-38050, Povo (Trento), Italy\\
            $^3$ Universit\'e Joseph Fourier c/o Institut de Biologie Structurale and
            CRG-IN13 at Institut Laue-Langevin (ILL), B.P. 156 Grenoble C\'edex 9, France.\\
            $^4$INFM\;CRS-SOFT, \,c/o \;
            Universit\`a \;di \,Roma \,``La \,Sapienza",I-00185, \,Roma, \,Italy
         }

\date{\today}

\begin{abstract}

We report new inelastic Raman and neutron scattering spectra for
glasses with different degree of fragility, v-SiO$_2$ , v-GeO$_2$,
(AgI)$_{0.5}$(Ag$_2$O-B$_2$O$_3$)$_{0.5}$,
(AgI)$_x$(AgPO$_3$)$_{1-x}$; the data are  compared for each
sample to obtain the Raman coupling function $C(\omega)$. The
study indicates a general linear behaviour of the $C(\omega)$ near
the Boson peak maximum, and evidence a correlation between
vibrational and relaxational properties, already observed in
recent publications.
\end{abstract}
 \maketitle
\section*{I. Introduction}
The evaluation of the vibrational density of states $g(\o)$ is
crucial for the study of the vibrational anomalies present in
disordered solids, among which we may mention the boson peak (BP)
\cite{Buc86,Buc88,Font90}, and the quasi-elastic scattering (QES)
\cite{sok93,Brod94}. In spite of a relevant effort of experimental
and theoretical work in recent years \cite{PM99,PM02,PM03}, many
questions about the origin of these universal features are still
unanswered. From Raman spectra alone, it is not possible to obtain
directly the density of states, because the expression of the
Raman intensity, $I_R$, besides $g(\o)$ contains a
frequency-dependent, and a-priori unknown, function $C(\o)$
\cite{Gal78}. $C(\o)$ measures the average scattering efficiency
of the vibrational modes with frequency lying between $\o$ and
$\o+d\o$. For Stokes scattering we have:
    \be
    I_R(\o,T) \propto \frac{n(\o,T)+1}{\o}g(\o)C(\o)
    \ee
where $n(\o,T)$ is the Bose-Einstein population factor. In
principle, not even neutron scattering does supply the true
$g(\o)$, because of the different scattering amplitudes of the
atomic components ; however, it has been demonstrated that in
glasses \cite{carpenter,Price87}, using the incoherent
approximation, the inelastic neutron scattering intensity, $I_N$,
is connected to the vibrational density of states  by an equation
similar to Eq. (1), but which does not contain any unknown
quantity such as $C(\o)$:
    \be
    I_N(\o,T) \propto \frac{n(\o,T)+1}{\o}g(\o)
    \ee
Hence, using equations (1) and (2) it is possible to evaluate the
most reliable $C(\o)$ and  assess the validity of the proposed
models.

Actually, different models have been proposed for the frequency
dependence of $C(w)$. In particular, for slightly distorted
plane-wave modes, $C(\o)$ has an $\o^2$ low-frequency dependence
\cite{Wint75,Nem77,MartGal81}; within the framework of the soft
potential model \cite{Karpov83,Karpov85} $C(\o)$ has a constant
value; finally, in the fracton-like model and at low frequency,
$C(\o)$ follows a power low, whose exponent depends both on the
model-system under investigation, and on the  mechanism of
polarizability modulation \cite{duv, font97, Viliani95}. Though
experimental efforts have been devoted to this problem,
\cite{Font99, sok02}, a definite conclusion as to the shape of
$C(\o)$ has not been reached, mainly due to the presence of
luminescence background and QES.

The aim of the present work is to determine the spectral shape of
$C(\o)$ by combined Raman and neutron scattering experiments, in
several kinds of glasses exhibiting different degrees of
fragility, i.e. v-SiO$_2$, v-GeO$_2$, and phosphate and borate
glasses.

\section*{ II. EXPERIMENTAL}
\subsection{ Sample preparation}

The v-SiO$_2$ sample was a commercial-grade spectrosil sample,
purchased from SILO (Florence). It had a disk shape with a
diameter of about 50 mm and a thickness of 4.8 mm. We have
prepared v-GeO$_2$, using reagent-grade GeO$_2$ powder (Aldrich
99.99\%). The powder was melted in ceramic crucibles for about 5h
at about 1600 C. The homogeneous and bubble-free melt was
subsequently quenched in air. The containers were cut and peeled
from the samples, leaving a cylinder-shaped bulk (15 mm diameter
and 5 mm thickness) of  clear and transparent glass.
Phosphate-glasses (AgI)$_x$(AgPO$_3$)$_{1-x}$ (from $x$=0 to
$x$=0.55) were prepared following the standard procedure
\cite{Minami79}. The powders were heated up to 500 C, the melted
solution was quenched and finally pressed onto a steel mask kept
at room temperature, in order to obtain disk-shaped samples with a
diameter of about 50 mm and a thickness of 2 mm. Also the borate
glass (AgI)$_{0.4}$-(Ag$_2$O-B$_2$O$_3$)$_{0.6}$ was prepared
following a standard procedure \cite{Magis79}  using special grade
AgI and AgNO$_3$ from Erba. For the B$_2$O$_3$ component, the
$^{11}$B-enriched material was supplied by Centronic Limited and
certified 97\% in $^{11}$B. In order to allow for a complete
reduction of AgNO$_3$ to Ag$_2$O, the powders were first heated in
a quartz crucible at 400C for 2 hours. Melting was achieved by
further heating at 500C for 5 hours. Quenching was obtained by
rapidly pouring and pressing the melted compound into circular
steel mould kept at 100C. The samples were then annealed at T$<$Tg
in order to minimize thermal stresses; they  have a disk shape
characterized by a diameter of 50 mm and a thickness of 2 mm. The
vitreous nature of melt quenched samples was verified by means of
X-ray diffraction measurements. The samples were subsequently
polished in order to obtain good optical surfaces. The fragility
of silica and germania is $m$=20, for phosphate and borate glasses
$m$=28, for polybutadiene $m$=60.
    \subsection{Neutron experiments}
The inelastic neutron scattering data were collected by using
different time-of-flight spectrometers at the Institut
Laue-Langevin (Grenoble, France). In particular, the IN6
spectrometer was used for phosphate and borate glasses, while IN4
was used for v-SiO$_2$ and v-GeO$_2$. The incident neutron
wavelength used was different according to the instrument used.
The time-of-flight data underwent the usual corrections, such as
subtraction of the empty cell and normalization to a vanadium
measurement. The density of states $g(\o)$ was obtained by an
iterative procedure described in the literature \cite{Cic99} and
corrected for the Debye-Waller factor and multiphonon
contributions, even if at 50 K where we do the Raman to Neutron
scattering ratio the multiphonon correction does not affect the
data significantly. More details about the procedure for the
treatment of data are reported elsewhere \cite{font04, fabiani04}.
    \subsection{Raman experiments}
Raman scattering experiments in HV geometry were performed using a
standard experimental set-up in a wide range of frequency (-300 to
5000 cm$^{-1}$). Such a wide range was necessary in order to take
properly into account the shape and temperature dependence of the
luminescence background, which is weak at room temperature, but
becomes important for a quantitative determination of the coupling
function at low frequency ($<$ 20cm-1) and low temperature ($<$
70K) Details on the data analysis are reported elsewhere
\cite{font04}. In addition, the low-frequency Raman spectra
contain also the QES. Luminescence and (especially) QES, cause a
great uncertainty in the determination of  the $C(\o)g(\o)$ term
in Eq. 1 from experimental data, and it is necessary to minimize
their effect. Accordingly, the experiments were performed at low
temperature (50 or 77 K for strong glasses), where the QES
intensity is negligible.
\begin{figure}[htb]
\includegraphics[width=9cm]{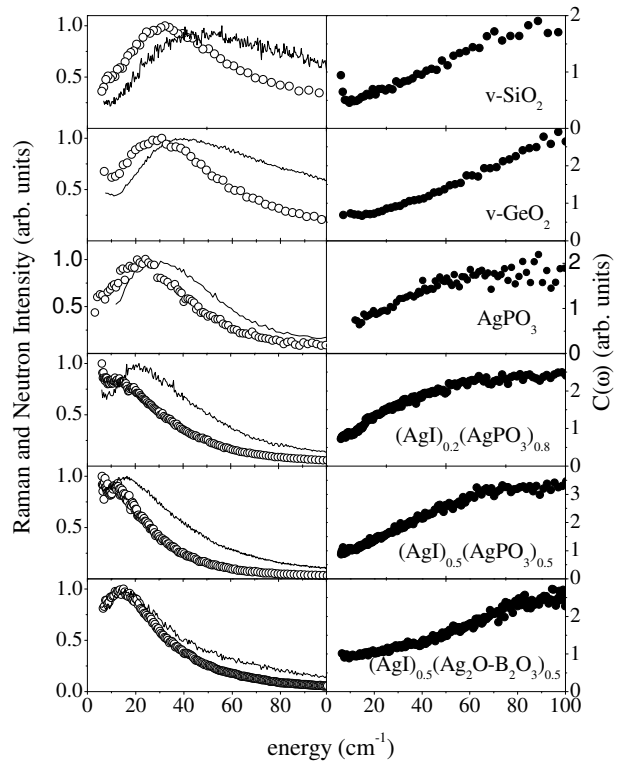}
\caption{Left panels: low frequency HV Raman (continuous lines)
and neutron scattering (open circles) spectra for the melt
quenched systems. Right panels: proportional to $C(\o)$,
calculated as the ratio between the Raman and neutron intensities.
For borate glasses $T=150$ K, for all other systems $T=50$ K.}
    \label{fig1}
\end{figure}
    \section*{III.Results and discussion}

In the left panels of Figure ~\ref{fig1}, we report the
(arbitrarily normalized) low-frequency Raman and neutron
scattering spectra for the melt-quenched samples at the
temperatures indicated in the caption, while in the right panels
of the same figure we report a quantity proportional to $C(\o)$ of
the same samples as calculated by the ratio between Raman and
neutron data. For all samples, Raman and neutron spectra have a
qualitatively similar shape, at all investigated temperatures.

\begin{figure}[htb]
\includegraphics[width=9cm]{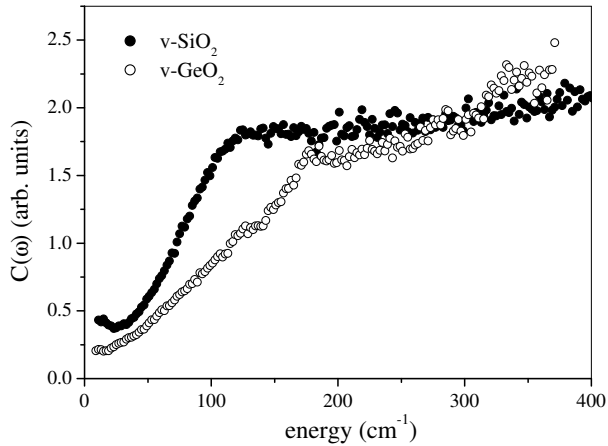}
\caption{ Room temperature $C(\o)$  for v-SiO$_{2}$ and
v-GeO$_{2}$ in an extended range.}
    \label{fig2}
\end{figure}

\begin{figure}[htb]
\includegraphics[width=9cm]{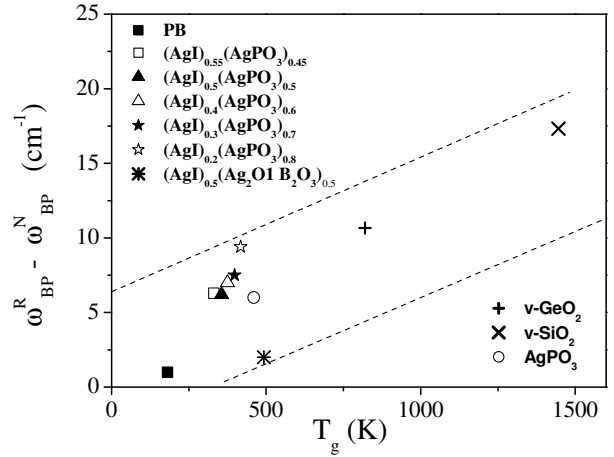}
\caption{ Energy difference between the maximum of the BP measured
by Raman and Neutron scattering, as a function of the glass
transition temperature $T_{g}$ of several glasses. Data on PB: A.
Brodin, private communicaton}
    \label{fig3}
\end{figure}

All coupling functions $C(\o)$ tend to a constant value at
frequency higher than the BP, as evidenced in Fig. ~\ref{fig2}
where we report, as examples, the cases of Si$O_{2}$ and Ge$O_{2}$
in a more extended frequency range; in particular, SiO$_{2}$
exhibits a linear behavior of $C(\o)$ in the BP frequency range,
followed by a nearly constant plateau up to high frequency where
the molecular modes begin to appear \cite{ font04, marioPC81}.
From Fig.~\ref{fig1} it appears that the maxima of the BP as
observed in Raman spectra are at higher energies with respect to
neutron spectra, an effect which is reflected in (or is a
consequence of) the linear behaviour of $C(\o)$ in the
corresponding left panels. This blue shift is larger in
v-SiO$_{2}$ and v-GeO$_{2}$ than in the other systems; in general,
the blue-shift is found to depend on the glass-transition
temperature $T_{g}$ of the system, as shown in Fig. ~\ref{fig3}
for several systems. Even if the scattering of data is large, the
trend appears to be significant. At first sight, the existence of
a such a correlation appears surprising, because the shift is
determined by vibrational properties (the density of vibrational
states and the Raman coupling function) which, in turn, depend on
the properties of the potential energy surface in the minima,
while $T_g$ is set by the distribution of barrier heights
separating adjacent minima (or basins of minima), and by the
geometry of the energy landscape in the neighbourhood of saddle
points. However, correlations between vibrational and relaxational
characteristics have already been observed and reported in recent
papers \cite{scopigno,nov,Bordat)}, and their origin is far from
clear. The present data, together with previously published ones,
seem to indicate that this behavuour is more common than might be
expected.
\section{Acknowledgments}
We would like to thank R. Dal Maschio for help in preparing the
v-GeO$_2$ samples; A. Brodin for communicating the shift of
polybutadiene; A. Bartolotta and G. Di Marco for the fragility
data of borate and phosphate glasses.

\end{document}